\documentclass[12pt,a4paper]{article}
\usepackage{amsmath}
\usepackage[dvips]{graphicx}
\input epsf
\usepackage{times}
\begin{document}
\begin{titlepage}
\title{Unitarity constraints  and role of geometrical
 effects in deep--inelastic scattering and  vector--meson electroproduction}
\author{S. M. Troshin and
 N. E. Tyurin\\[1ex]
\small  \it Institute for High Energy Physics,\\
\small  \it Protvino, Moscow Region, 142280, Russia}
\normalsize
\date{}
\maketitle

\begin{abstract}
Deep--inelastic scattering at low $x$ and elastic  vector meson
electroproduction are analyzed  on the basis
of the $s$--channel unitarity extended to
off--shell particle scattering. It appeared that the
role of  unitarity is important but contrary to the
case of the on-shell  scattering it does not rule out
power--like behavior of the total cross--sections.
We discuss  behavior of  the total cross--section of virtual
photon--proton scattering in the geometrical approach
and obtain  that the exponent of the power-like
energy  dependence of $\sigma^{tot}_{\gamma^* p}$ is related
to the constituent quark interaction radius.  The  mass effects
and  energy dependence of vector meson electroproduction
are discussed alongside with the angular
distributions at large momentum transfers in these processes.
 \\[2ex]
\end{abstract}
\end{titlepage}
\setcounter{page}{2}

\section*{Introduction}
 Rising
dependence of
 the virtual--photon proton scattering total cross--section
on the center of mass energy $W^2$ discovered at HERA  \cite{her}
 led to the renewed interest in the  mechanism of
 diffraction at high energies.
Such behavior was in fact  predicted in \cite{indur} and expected in
perturbative QCD \cite{pqcd}.
The total virtual photon--proton cross--section is related
to the structure function $F_2$  at small $x$.
The HERA effect
 is consistent with various $W^2$ -- dependencies of $\sigma^{tot}_{\gamma^*p}$
  and has been explained in the different ways,
  among them is a manifestation of
 hard BFKL Pomeron \cite{lipa}, an appearance  of the DGLAP evolution
 in  perturbative QCD \cite{pqcd},
 a transient phenomena, i.e. preasymptotic effects \cite{nad}
 or a true asymptotical off--mass--shell scattering amplitude
behavior \cite{petr}. There is an extensive list of papers
devoted to this subject and many interesting
results are described in the review papers
 (cf. e.g. \cite{her,lands}).
The
strong  rise of the structure function $F_2$ at small $x$
 which can be described by the power-like dependence\footnote{The
  most recent results of H1 Collaboration \cite{h1}
confirmed an $x$-independence of the exponent $\lambda$ at low $x$
by measuring
the derivative $\left( \frac{\partial \ln F_2(x,Q^2)}{\partial
\ln x}\right)_{Q^2}$ as a function both of $Q^2$ and of $x$ for the first time.
Some provisions against this independence were mentioned in \cite{mart},
moreover there are also other parameterizations which  describe
 the experimental data equally well  (cf. e.g. \cite{wolf}).
 Despite that, it seems
 that the parameterization (\ref{wlq}) provides the most natural way to approximate
 the available  experimental data.} $F_2(x,Q^2)\propto x^{-\lambda(Q^2)}$ implies that
\begin{equation}\label{wlq}
\sigma^{tot}_{\gamma^*p}(W^2,Q^2)\propto (W^2)^{\lambda(Q^2)}
\end{equation}
with $\lambda(Q^2)$ rising with $Q^2$ from about $0.1$ to
about $0.5$ and
is  regarded as a somewhat  surprising fact.
It is  due to the fact that according to the experimental data an
 energy dependence of the total cross--sections in hadronic
 interactions, the total cross--section increase is rather slow
  ($\lambda\sim 0.1$).
The mentioned variance may be regarded as not a fundamental one. First,
 there is no
Froissart--Martin bound for the case
off--shell particles \cite{indur,petr}.  Additional
assumptions are needed to reinstate this bound \cite{ttpre,levin}.
Second,  it cannot be granted that
the  preasymptotic effects and approach to the asymptotics
are the same for the on--shell and off--shell scattering.
It seems that for  some  reasons scattering of virtual
particles reaches the  asymptotics faster than the scattering of
the real particles.

It is worth noting
that the space-time structure of
the low--$x$ scattering
 involves  large distances
$l\sim 1/mx$ on the light--cone \cite{pas}, and the
region of $x\sim 0$ is  sensitive to the  nonperturbative contributions.
Deep--inelastic scattering in this region turns out to be a coherent
process where diffraction plays a major role and nonperturbative models
such as  Regge or vector dominance model can be competitive with perturbative
QCD and successfully applied for  description of the experimental data.

 It is essential to obey the general principles in the nonperturbative region
and, in particular, to satisfy unitarity.
The most common form of unitarity solution -- the eikonal one --
 was generalized for
the off--shell scattering in \cite{petr}.
In this paper we consider an
off--shell extension of the $U$--matrix approach to the amplitude
unitarization.  It is
shown that this approach along with
 the respective extension of the  quark model for the
 $U$--matrix \cite{csn}  leads to (\ref{wlq}), where
the exponent $\lambda (Q^2)$ is related to the $Q^2$--dependent
 interaction radius attributed to  constituent quark. These results
cannot be obtained in the eikonal unitarization  which
which reproduces bare ``Born'' input form with subleading corrections
 for the output amplitude in the case of the
 off--shell scattering \cite{petr}. The fundamental distinction between
 the two forms of amplitude unitarization is in the analytical properties
 in the complex energy plane \cite{blan}.

It is worth noting an  importance of the effective interaction
radius concept \cite{log}.  The study of the effective interaction radius
dependence on the scattering variables appeared to be  very useful for
understanding of the dynamics of high energy hadronic
reactions \cite{chy,khru}.
It is widely known nowadays  that the respective
geometrical considerations  provide a deep insight
in hadron dynamics and deep--inelastic scattering (cf. \cite{bart}).

Besides  the studies of deep--inelastic scattering (DIS) at low $x$
the measurements of the characteristics
of the elastic
vector meson (VM) production were performed in the experiments H1 and
ZEUS at HERA \cite{zosa,melld}. It was shown that the integral  cross section
of the elastic vector meson production increases with energy in the way
similar to the
$\sigma^{tot}_{\gamma^*p}(W^2, Q^2)$ dependence
 on  $W^2$ \cite{her}.
It appeared that an increase of   VM
electroproduction cross--section  with energy is  steeper for
heavier vector mesons as well as  when the virtuality
$Q^2$ is higher. Discussion of such a behavior in  various model
approaches based on the nonperturbative hadron physics or
perturbative QCD can be found in (cf. e.g.   \cite{lands}).

   Application of approach based on the
off-shell extension of the $s$--channel unitarity
to elastic vector meson electroproduction
$\gamma^*p\to Vp$ allows  to obtain angular
dependence and predict interesting mass effects in these processes.
 It appears that
the obtained mass and $Q^2$ dependencies do not contradict
to the experimentally observed trends.
It is also valid for the angular
distributions at large momentum transfers.

\section{Off--shell unitarity}
Extension of the $U$--matrix unitarization scheme for the off-shell
scattering was considered in brief in \cite{ttpre}.
Here we give a more detailed treatment of this problem.
We adopt a commonly accepted picture of DIS at small $x$, i.e.
it is supposed that
the virtual  photon fluctuates into a quark--antiquark
pair $q \bar q$ and this pair  is considered as an effective
 virtual  vector meson state in the processes with small  $x$.
This effective virtual meson  interacts then with a hadron. For simplicity
we consider  single effective vector meson field. We use
for the amplitudes  of the processes
\begin{equation}
V^*+h  \to  V^*+h,\quad
V^*+h \to  V+h\quad \mbox{and}\quad
V+h \to  V+h
\end{equation}
the notations $F^{**}(s,t,Q^2)$, $F^{*}(s,t,Q^2)$ and $F(s,t)$, respectively,
i. e. we denoted in that way the
amplitudes when both initial and final mesons  are off mass
shell, only initial meson is off mass shell and both mesons are on
mass shell.

The unitarity relation  for the amplitudes $F^{**}$ and
$F^*$ has a similar structure as the unitarity equation for
the  on--shell
amplitude $F$ but relates, in fact,  different amplitudes.
Therefore, the  unitarity constraints in DIS are much less stringent
than in hadron--hadron scattering.
In  impact
parameter representation at high energies it  relates the
amplitudes $F^{**}$ and $F^*$ in the following way
\begin{equation}\label{offs}
\mbox{Im} F^{**}(s,b,Q^2) =  |F^{*}(s,b,Q^2)|^2+\eta^{**}(s,b,Q^2),
\end{equation}
where $\eta^{**}(s,b,Q^2)$ is the contribution to the unitarity of
many--particle intermediate on--shell states. The function
$\eta^{**}(s,b,Q^2)$ is the sum of the $n$--particle production
cross--section in the process of the virtual meson interaction
with a hadron $h$, i. e.
\[
\eta^{**}(s,b,Q^2)=\sum_n \sigma_n(s,b,Q^2).
\]
There is a similar relation  for the functions $F^{*}$ and $F$,
\begin{equation}\label{offs1}
\mbox{Im} F^{*}(s,b,Q^2) =  F^{*}(s,b,Q^2)F(s,b,Q^2)+\eta^{*}(s,b,Q^2).
\end{equation}
Contrary to $\eta^{**}(s,b,Q^2)$
the function $\eta^{*}(s,b,Q^2)$ has no simple physical
 meaning and it will be discussed later.
\begin{figure}[htb]
\vspace{-0.9in}
\hspace{1.5cm}
\epsfxsize=4.5in \epsfysize=2.1in
 \epsffile{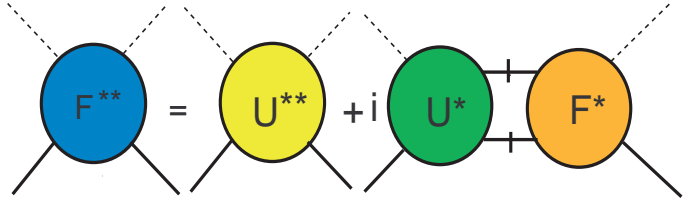}
 \caption[illi]{The solution of the off--shell unitarity relation for the
 amplitude $F^{**}$.}
\label{ill1}
\end{figure}
\begin{figure}[htb]
\vspace{-0.9in}
\hspace{1.5cm}
\epsfxsize=4.5in \epsfysize=2.1in
 \epsffile{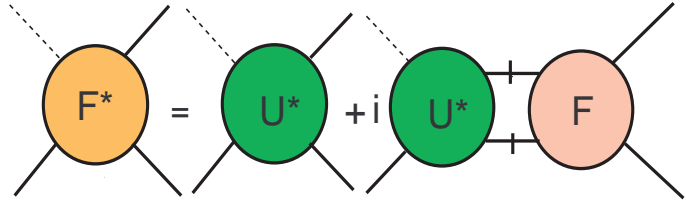}
 \caption[illii]{The solution of the off--shell unitarity relation for the
 amplitude $F^{*}$.}
\label{ill2}
\end{figure}
The solution of the off--shell unitarity relations can be graphically
represented   for the amplitudes $F^{**}$ and $F^{*}$ in the Figs. \ref{ill1}
 and \ref{ill2}
respectively and
 has a simple form in the impact
parameter representation :
\begin{eqnarray}
F^{**}(s,b,Q^2) & = & U^{**}(s,b,Q^2)+iU^{*}(s,b,Q^2)F^{*}(s,b,Q^2)\nonumber\\
F^{*}(s,b,Q^2) & = & U^{*}(s,b,Q^2)+iU^{*}(s,b,Q^2)F^{}(s,b).\label{es}
\end{eqnarray}
 It is worth noting that the solution of the off--shell unitarity
in the nonrelativistic case for a $K$--matrix representation
was obtained for the first time in \cite{lov}.
The solution  of this system is the following
\begin{equation}\label{fv}
F^{*}(s,b,Q^2)  =  \frac{U^{*}(s,b,Q^2)}{1-iU(s,b)}=
\frac{U^{*}(s,b,Q^2)}{U(s,b)}F(s,b)
\end{equation}
\[
F^{**}(s,b,Q^2)  =  \frac{U^{**}(s,b,Q^2)}{1-iU(s,b)} -
i\frac{U^{**}(s,b,Q^2)U(s,b)-[U^{*}(s,b,Q^2)]^2}{1-iU(s,b)}=
\]
\begin{equation}
\frac{U^{**}(s,b,Q^2)}{U(s,b)}F(s,b) -
iU^*(s,b,Q^2)[\frac{U^{**}(s,b,Q^2)}{U^*(s,b,Q^2)}
-\frac{U^{*}(s,b,Q^2)}{U(s,b)}]F(s,b),
\label{vrss}
\end{equation}
where the on--shell amplitude $F(s,b)$ has the following representation
\begin{equation}
F(s,b)={U(s,b)}/{[1-iU(s,b)]}
\end{equation}
We assume the following relation to be valid
at the  level of the ``Born'' amplitudes in the impact parameter
space
\begin{equation}
\frac{U^{*}}{U}=\frac{U^{**}}{U^*}.\label{zr}
\end{equation}
This relation is
valid, e. g. in the Regge model with factorizable residues
and the $Q^2$--independent trajectory.
It is also valid in the
off--shell extension of the chiral quark model for the $U$--matrix
which we will consider further.
Eq. (\ref{zr}) implies the following forms for  the impact parameter
dependent functions $U^{*}$ and $U^{**}$:
\begin{eqnarray}
U^{*}(s,b,Q^2) & = & \omega (s,b,Q^2)U(s,b)\nonumber\\
U^{**}(s,b,Q^2) & = & \omega(s,b,Q^2)U(s,b)\omega(s,b,Q^2).\label{fct}
\end{eqnarray}
This factorization may be treated as a reflection of the
 universality of the initial and final state
interactions responsible for the transitions between the on and off mass
 shell states.
It seems to be a quite natural assumption.
Note, that this factorization does not
survive  for the
 amplitudes $F(s,t)$, $F^*(s,t,Q^2)$
 and $F^{**}(s,t,Q^2)$, i.e.  after Fourier-Bessel
  transform is performed,

 Thus, we  have for the amplitudes $F^*$ and $F^{**}$
 the following relations
\begin{eqnarray}
F^{*}(s,b,Q^2) & = & \frac{U^{*}(s,b,Q^2)}{1-iU(s,b)}=
\omega(s,b,Q^2)F(s,b)
\label{vrq}\\
F^{**}(s,b,Q^2) & = & \frac{U^{**}(s,b,Q^2)}{1-iU(s,b)}=
\omega(s,b,Q^2)F(s,b)\omega(s,b,Q^2)\label{vr}
\end{eqnarray}
and unitarity  provides inequalities
\begin{equation}
|F^*(s,b,Q^2)|  \leq  |\omega (s,b,Q^2)|,\quad
|F^{**}(s,b,Q^2)|  \leq  |\omega^2(s,b,Q^2)|.\label{bnd}
\end{equation}
It is worth noting that the above limitations are   much less stringent
than the limitation for the on--shell amplitude $|F(s,b)|\leq 1$.
As a result, there is no Froissart--Martin bound in DIS at low $x$ and
experimentally observed power-like energy dependence of the total cross--section
can represent a true asymptotical dependence.

 When  the function $\omega(s,b,Q^2)$ is real  we can write down
a simple expression for the inelastic overlap function
$\eta^{**}(s,b,Q^2)$:
\begin{equation}\label{etaf}
  \eta^{**}(s,b,Q^2)  =
\omega(s,b,Q^2)\frac{\mbox{Im} U(s,b)}{|1-iU(s,b)|^2}\omega(s,b,Q^2)
\end{equation}
The following relation is  valid for the function
$\eta^{*}(s,b,Q^2)$:
\begin{equation}\label{etazv}
  \eta^{*}(s,b,Q^2)=[\eta^{**}(s,b,Q^2)\eta^{}(s,b)]^{1/2}.
\end{equation}
Eq. (\ref{etazv}) allows one to connect the integral
\[
\Sigma (s,Q^2)\equiv 8\pi\int_0^\infty\eta^{*}(s,b,Q^2)bdb
\]
with the total inelastic cross--section:
\[
\Sigma (s,Q^2)|_{Q^2\to 0}=\sigma_{inel}(s).
\]

\section{Off--shell  scattering  and the quark model for $U$--matrix}
The above formulas are rather general and  are  not
useful alone under analysis of the data.
We need an explicit form for the functions $U$, $U^*$ and $U^{**}$
and therefore a phenomenological model is to be constructed.
As a starting point
we use a  quark model for the hadron scattering
 described in \cite{csn}.
In this section we list the main features  and then
construct an  off--shell extension of the  model.
In fact it is  based on the ideas of chiral quark models.
The  picture of hadron structure in the model
 with the valence constituent quarks
located in the central part and the surrounding condensate
implies  that     the
overlapping of hadron structures and interaction of the
condensates
 occur  at  the first stage of  collision and results in  generation
of the quasiparticles, i.e. massive quarks  (cf. Fig. 3).
\begin{figure}[htb]
\hspace{2.5cm}
\epsfxsize=3in \epsfysize=1.85in\epsffile{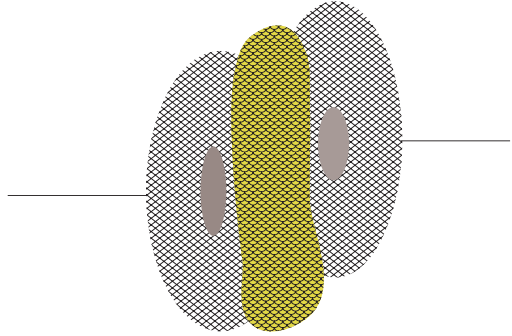}
 \caption[illyi]{Schematic view of initial stage of the hadron
 interaction and formation of the effective field.}
\label{ill5}
\end{figure}

 These quarks play role
of scatterers.
 To estimate number
of such quarks one could assume that  part of hadron energy carried by
the outer condensate clouds is being released in the overlap region
 to generate massive quarks. Then their number can be estimated  by
the quantity:
 \begin{equation} \tilde{N}(s,b)\,\propto
\,\frac{(1-k_Q)\sqrt{s}}{m_Q}\;D^{h}_c\otimes D^{V}_c,
\label{4}
\end{equation} where $m_Q$ -- constituent quark mass, $k_Q$ --  fraction
hadron energy carried  by  the constituent valence quarks. Function $D^h_c$
describes condensate distribution inside the hadron $h$, and $b$ is
an impact parameter of the colliding hadron $h$ and meson $V$.
Thus, $\tilde{N}(s,b)$ quarks appear in addition to $N=n_h+n_V$
valence quarks. Those quarks are transient
ones: they are transformed back into the condensates of the final
hadrons in elastic scattering. It should be noted that we use subscript
$Q$ to refer  the  constituent quark $Q$ and the same letter $Q$
is used to denote a virtuality $Q^2$. However, they enter formulas
in a way excluding  confusion.

\begin{figure}[htb]
\hspace{3.9cm}
\epsfxsize=2in \epsfysize=1.5in
 \epsffile{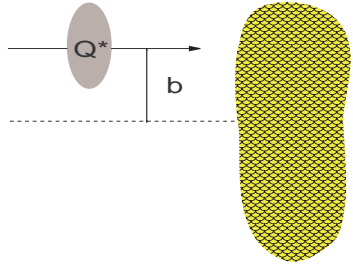}
 \caption[illy]{Schematic view of the virtual constituent quark $Q^*$
 scattering in the effective field generated by $N_{sc}(s,b)$ scatterers, where
 $N_{sc}(s,b)=\tilde N(s,b)+N-1$.}
\label{ill4}
\end{figure}

In the model  the valence quarks located in the
central part of a hadron are supposed to scatter in a
quasi-independent way by the effective field.
Due to quasi--independence of the valence quarks scatterings the basic
dynamical  quantity (the function $U$) can be factorized.
When  one of the hadrons
(vector meson in our case)  is off mass shell,  the corresponding
function $U^{**}(s,b,Q^2)$ is represented as the
following product
\begin{equation} U^{**}(s,b,Q^2)\,=\, \prod^{n_{h}}_{i=1}\,
\langle f_{Q_i}(s,b) \rangle \prod^{n_{V}}_{j=1}\, \langle
f^{}_{Q^*_j}(s,b,Q^2) \rangle .\label{prdv} \end{equation}
The factors $\langle
f_{Q}(s,b)\rangle$ and $\langle f_{Q^*}(s,b,Q^2)\rangle$
 correspond to the   individual quark
scattering amplitudes smeared over  constituent
 quark transverse position  and
the fraction of longitudinal
momentum carried by this quark.
Under the virtual constituent quarks $Q^*$ we mean the ones
 composing the virtual
 meson.
Factorization (\ref{prdv}) reflects the coherence in the valence
quark scattering, i.e. all valence quarks are scattered in the
effective field simultaneously and there are no spectator valence quarks.
This factorization might  be considered as an effective
implementation of constituent quarks' confinement.
The averaged amplitudes $\langle f_Q(s,b)\rangle $ and
$\langle f_{Q^*}(s,b,Q^2)\rangle $ describe elastic
scattering  of a single
valence  on-shell   or off--shell   quarks $Q$ and $Q^*$, respectively,
off the effective field (cf. Fig. 4). We use for the function $\langle
f_Q(s,b)\rangle $  the following expression
\begin{equation}
\langle f_Q(s,b)\rangle =[\tilde{N}(s,b)+(N-1)] \,V_Q(\,b\,)
\label{7}
\end{equation}
where $V_Q(b)$  has a simple form
\[
V_Q(b)\propto g\exp(-m_Qb/\xi ),
\]
which corresponds to quark interaction radius
\[
r_Q=\xi/m_Q.
\]
The function $\langle f_Q^*(s,b,Q^2)\rangle $ is to be written
as
\begin{equation}
\langle f^{}_{Q^*}(s,b,Q^2)\rangle =[\tilde{N}(s,b)+(N-1)] \,V_{Q^*}(b, Q^2).
\label{fqv}
\end{equation}
In the above equation
\begin{equation}
V_{Q^*}(b,Q^2)\propto g(Q^2)\exp(-m_Qb/\xi (Q^2) )
\end{equation}
and this form corresponds to the virtual constituent quark
interaction radius
\begin{equation}\label{rqvi}
r_{Q^*}=\xi(Q^2)/m_Q.
\end{equation}
The functions $V_{Q}(b)$ and $V_{Q^*}(b,Q^2)$ in the model are associated
with the "matter" distribution inside  constituent quarks and can be
considered as  strong formfactors.

Equations (\ref{7},\ref{fqv}) imply that each valence quark is being scattered
by all other $N-1$ valence quarks belonging to the same hadron as well as
to the other hadron and by $\tilde{N}(s,b)$ quarks produced by the
excitation of the chiral condensates.
Due to the different radii the $b$--dependence of $\tilde{N}(s,b)$
being weak compared to the
$b$--dependence of $V_Q$ or $V_{Q^*}$ and
this function can be approximately taken to be
 independent on the impact parameter $b$.
Dependence on virtuality $Q^2$ comes through dependence of the
intensity of the virtual constituent
 quark interaction $g(Q^2)$
and the $\xi(Q^2)$, which determines the quark
interaction radius (in the on-shell limit $g(Q^2)\to g$ and
$\xi(Q^2)\to\xi$).

Introduction of the $Q^2$ dependence into the interaction radius of a constituent
quark which in the present approach consists of a current quark
and the  cloud of quark--antiquark pairs of the different
flavors is the main issue of the off--shell extension of the
model and the origin of this
dependence and its possible physical interpretation will be discussed
in Section 6.

According to these considerations the explicit functional forms
for the generalized
reaction matrices $U^*$ and $U^{**}$
 can easily be written in the form of (\ref{fct}) with
\begin{equation}\label{omeg}
  \omega(s,b,Q^2)=\frac{\langle f^{}_{Q^*}(s,b,Q^2)\rangle}
  {\langle f^{}_{Q}(s,b)\rangle}.
\end{equation}
Note that  (\ref{zr}) and (\ref{fct}) imply that the amplitude of
the process $Q^*\to Q$ is the following
\[
\langle f^{}_{Q^*\to Q}(s,b,Q^2)\rangle=
[\langle f^{}_{Q^*}(s,b,Q^2)\rangle \langle f^{}_{Q}(s,b)\rangle]^{1/2}.
\]

We consider the high--energy limit and for the simplicity assume here
that all the constituent quarks have equal masses and parameters
$g$ and $\xi$ as well as $g(Q^2)$ and $\xi(Q^2)$
do not depend on quark flavor.
 We also assume
pure imaginary amplitudes. Then  the functions
$U$, $U^*$ and $U^{**}$ are g
\begin{equation}
 U(s,b)  =  ig^N\left (\frac{s}{m^2_Q}\right )^{N/2}
\exp \left [-\frac{m_QNb}{\xi}\right ] \label{usb}
\end{equation}
\begin{equation}
U^*(s,b,Q^2)  =  \omega (b,Q^2) U(s,b),\quad
 U^{**}(s,b,Q^2)  =  \omega^2 (b,Q^2) U(s,b)\label{uvv}
\end{equation}
where the function $\omega$ is an energy-independent one
and has the following dependence on $b$ and $Q^2$
\begin{equation}\label{ome}
  \omega(b,Q^2)  =
  \frac{g(Q^2)}{g}\exp \left [-\frac{m_Qb}{\bar{\xi}(Q^2)}\right ]
\end{equation}
with
\begin{equation}\label{ksi}
  \bar{\xi}(Q^2)=\frac{\xi\xi(Q^2)}{\xi-\xi(Q^2)}.
\end{equation}

\section{Total cross--sections of $\gamma^* p$ interactions}

With explicit forms of the functions $U$, $U^*$ and $U^{**}$ the corresponding
scattering amplitudes can be calculated. The most simple case is the forward
scattering at large energies.
For the on--shell scattering when $\omega \to 1$
at large $W^2$, the
total photoproduction cross--section  has a Froissart-like asymptotic behavior
\begin{equation}\label{ons}
  \sigma^{tot}_{\gamma p}(W^2)\propto\frac{\xi^2}{m_Q^2}\ln ^2 \frac{W^2}{m_Q^2},
\end{equation}
where the usual for DIS notation $W^2$
 instead of $s$ is used. Similar result is valid also for the off mass shell
 particles if the interaction radius of virtual quark does not depend
 on $Q^2$ and is equal to the interaction radius of the on--shell quark,
 i.e. $\xi(Q^2)\equiv \xi $. The behavior of the total cross--section
 at large $W^2$
\begin{equation}\label{ofs}
  \sigma^{tot}_{\gamma^* p}(W^2)\propto
  \left[\frac{g(Q^2)\xi}{gm_Q}\right]^2
  \ln ^2 \frac{W^2}{m_Q^2}.
\end{equation}

We consider next the off-shell scattering and suppose now
that $\xi(Q^2)\neq\xi$.
It should be noted
first that for the case when $\xi(Q^2)<\xi$
the total   cross--section would be energy-independent
\[
\sigma^{tot}_{\gamma^* p}(W^2)\propto C(Q^2)\equiv\left[\frac{g(Q^2)\xi}
{g\lambda(Q^2)m_Q}\right]^2\]
 in the asymptotic region.
This
 scenario would  mean that the experimentally observed rise of
  $\sigma^{tot}_{\gamma^* p}$
 is transient preasymptotic phenomenon.
  It can be realized when we
 replace  the mass $m_Q$ by the quantity
   $m_{Q^*}=\sqrt{m_Q^2+Q^2}$ in order to obtain
  the interaction radius of the off-shell constituent quark and
  write it down as $r_{Q^*}=\xi/m_{Q^*}$, or equivalently
  replace  $\xi$ by
$\xi(Q^2)={\xi m_Q}/{\sqrt{m_Q^2+Q^2}}$.
The above
option cannot be excluded in principle, however, it is a self-consistent choice
in the framework of the model only at large $Q^2\gg m_Q^2$ since it was
originally supposed
that the function $\xi$ is universal for the different quark flavors.

However, when $\xi(Q^2)>\xi$ the situation is different
and we have at large
$W^2$
\begin{equation}\label{totv}
\sigma^{tot}_{\gamma^* p}(W^2,Q^2)\propto G(Q^2)\left(\frac{W^2}{m_Q^2}
\right)^{\lambda (Q^2)}
\ln \frac{W^2}{m_Q^2},
\end{equation}
where
\begin{equation}\label{lamb}
\lambda(Q^2)=\frac{\xi(Q^2)-\xi}{\xi(Q^2)}.
\end{equation}
We shall further concentrate  on this self-consistent for any
$Q^2$ values and
the most interesting case.

All the above expressions for $ \sigma^{tot}_{\gamma^* p}(W^2)$ can
be rewritten as the corresponding dependencies of $F_2(x,Q^2)$ at small $x$
according to the relation
\[
F_2(x,Q^2)=\frac{Q^2}{4\pi^2\alpha}
 \sigma^{tot}_{\gamma^* p}(W^2),
\]
where $x=Q^2/W^2$.

In particular,
(\ref{totv}) will appear in the  form
\begin{equation}\label{totv1}
F_2(x,Q^2)\propto\tilde{G}(Q^2)\left(\frac{1}{x}\right)^{\lambda (Q^2)}
\ln (1/x),
\end{equation}

It is interesting that the value and $Q^2$ dependence of the
 exponent $\lambda(Q^2)$ is related to the interaction radius
 of the virtual constituent quark. The value of parameter $\xi$
 in the model is determined by the slope of the differential cross--section
of elastic scattering at large $t$ \cite{lang}, i. e.
\begin{equation}\label{ore}
  \frac{d\sigma}{dt}\propto\exp\left[-\frac{2\pi\xi}{m_QN}\sqrt{-t}\right]
\end{equation}
and from the $pp$-experimental data it follows that $\xi=2$.
Then from the data for $\lambda(Q^2)$ obtained
at HERA \cite{h1} we can calculate the ``experimental'' $Q^2$--dependence
 of the function
$\xi(Q^2)$:
\begin{equation}\label{ksiq}
\xi(Q^2)=\frac{\xi}{1-\lambda(Q^2)}.
\end{equation}
 Evidently experiment
 indicates that $\xi(Q^2)$ rises with $Q^2$. This rise is slow and consistent with
 $\ln Q^2$ extrapolation (Fig. 5):
\[
\xi(Q^2)=\xi + a\ln\left(1+ \frac{Q^2}{Q_0^2}\right),
\]
where $a=0.172$ and $Q_0^2=0.265$ GeV$^2$.
  Assuming this dependence
for higher values of $Q^2$ and using (\ref{lamb}), we then
   predict saturation
 of $\lambda(Q^2)$ at large $Q^2$, i.e.  the
 flattening should  take place:
 \[
 \lambda (Q^2)={a\ln\left(1+ \frac{Q^2}{Q_0^2}\right)}
 /{\left[\xi + a\ln\left(1+ \frac{Q^2}{Q_0^2}\right)\right]}.
 \]
\begin{figure}[t]
\vspace{2mm}
\hspace{1.5cm}
\epsfxsize=4.1in \epsfysize=3.2in
 \epsffile{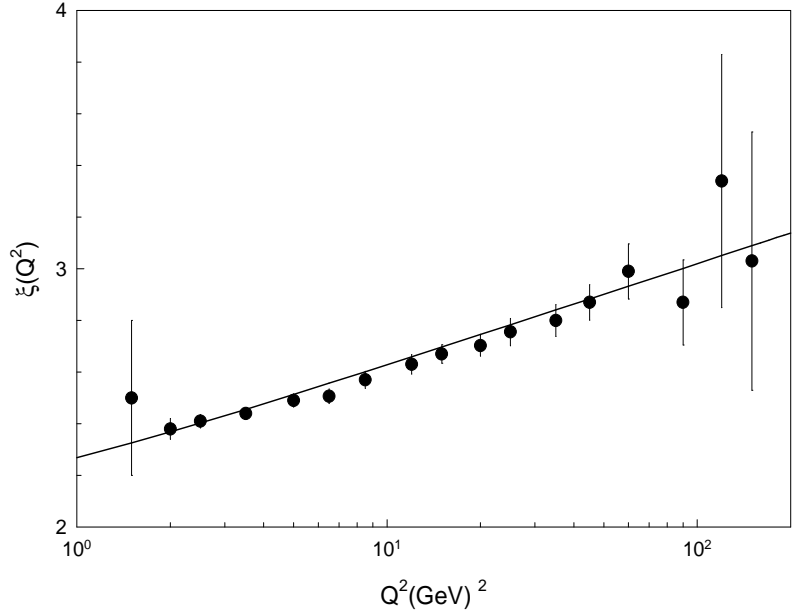}
 \caption[ksi]{ The ``experimental'' behavior of the function $\xi(Q^2)$.}
\label{fig:1}
\end{figure}
Increase of $\xi(Q^2)$  corresponds to the increasing interaction
 radius of  constituent
quarks from the virtual vector meson which is illustrated on Fig. 6.
\begin{figure}[htb]
\hspace{4.5cm} \epsfxsize=3in \epsfysize=1.5in
 \epsffile{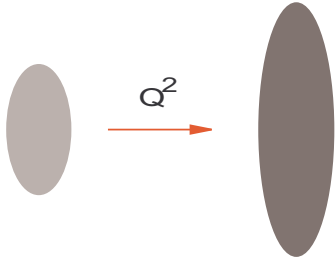}
 \caption[illiii]{ The increase with virtuality of the constituent quark interaction
 radius.}
\label{ill3}
\end{figure}

\section{Elastic vector meson production}
The calculation of the elastic and inelastic cross--sections can also be
directly performed similar to the calculation of the total cross-sections
using (\ref{usb}), (\ref{uvv}) and (\ref{ome}) and integrating
 the functions $|F^*(s,b,Q^2)|^2$
and $\eta^{**}(s,b,Q^2)$ over  impact parameter. The following
asymptotic dependencies for the cross--sections of elastic
 scattering and inelastic
interactions are obtained in that way
\begin{equation}\label{elcrs}
\sigma^{el}_{\gamma^* p}(W^2,Q^2)\propto G_e(Q^2)\left(\frac{W^2}{m_Q^2}
\right)^{\lambda (Q^2)}
\ln \frac{W^2}{m_Q^2}
\end{equation}
and
\begin{equation}\label{inelcrs}
\sigma^{inel}_{\gamma^* p}(W^2,Q^2)\propto G_i(Q^2)\left(\frac{W^2}{m_Q^2}
\right)^{\lambda (Q^2)}
\ln \frac{W^2}{m_Q^2}
\end{equation}
with the universal exponent $\lambda (Q^2)$ given by Eq.
(\ref{lamb}).

The above relations imply that the ratios of elastic and inelastic
cross--sections to the total one
do not depend on energy. In order to confront these results with
experimental data it is useful to keep in mind that as it was noted in \cite{pred},
diffraction of the virtual photon
includes both elastic and inelastic scattering
of its fluctuations.

Now we  consider elastic (exclusive) cross--sections
both for the light and heavy
vector mesons production.
We assumed earlier that the virtual
photon before the interaction with the proton fluctuates into the
$\bar Q Q$ -- pair and for simplicity we limited ourselves with  light
quarks under discussion of the total cross--section.
Therefore we need to get rid of the light
quark limitation and extend the above approach in order to
include  the quarks with the different masses.
The inclusion, in particular, of heavy vector meson production into this
scheme is straightforward: the virtual photon fluctuates before
the interaction with proton into the heavy quark--antiquark pair
 which constitutes
the virtual heavy vector meson state. After the interaction with a proton
this state becomes a real  heavy vector meson.

Integral exclusive (elastic) cross--section of vector meson production in
the process $\gamma^*p\to Vp$ when the  final state vector meson
 contains not only the  light quarks can be calculated directly
according to the above scheme and formulas of Section 2:
\begin{equation}\label{elvec}
\sigma^{V}_{\gamma^* p}(W^2,Q^2)\propto G_{V}(Q^2)\left(\frac{W^2}
{{m_Q}^2}
\right)^{\lambda_{V} (Q^2)}
\ln \frac{W^2}{{m_Q}^2},
\end{equation}
where
\begin{equation}\label{lavm}
\lambda_{V}(Q^2)= \lambda (Q^2)\frac{\tilde{m}_Q}{\langle m_Q \rangle}.
\end{equation}
In (\ref{lavm}) $\tilde{m}_Q$ denotes the mass of the constituent
quarks from
the vector meson and $\langle m_Q \rangle$ is the mean constituent
quark mass
 of the vector meson and proton system.
Evidently
$\lambda_{V}(Q^2)=\lambda(Q^2)$
for the light vector mesons.
In the case when the vector meson
is very heavy, i.e. $\tilde m_Q\gg m_Q$ we
have
\[
\lambda_{V}(Q^2)=\frac{5}{2}\lambda(Q^2).
\]
We conclude that the respective cross--section
rises faster than the corresponding cross--section
of the light vector meson production, e.g. (\ref{lavm}) results in
\[
\lambda_{J/\Psi}(Q^2)\simeq 2\lambda(Q^2).
\]
The above results are in a qualitative agreement with the
trends observed in the HERA experiments \cite{zosa,melld}.

\section{Angular structure of elastic
vector meson production}
Now we turn to calculation of the scattering amplitudes at $t\neq 0$.
It will allow us to get a differential cross-sections and to confront
the results with the
first measurements  of angular distributions at large $t$
in the light vector meson production \cite{zdta}.
 It was found that the angular
distribution in the proton--dissociative
 processes \cite{critt} is consistent
 with the power dependence $(-t)^{-3}$.
Calculation of the differential
cross--sections in elastic vector meson production can be performed
using the analysis
of the singularities of the amplitudes in the complex impact parameter plane
which was applied for elastic hadron scattering in \cite{lang}.
There are different approaches to the vector meson production, e.g.
 recent application
of the geometrical picture was given in \cite{cald}. Angular distributions
 can be described also in
the approaches based on the perturbative QCD \cite{ivan,forsh,gots} which provides
smooth power-like $t$-dependence. Brief review of the recent results of these
approaches can be found in \cite{diehl}.

Since the integration in the Fourier-Bessel transform
 goes over the variable $b^2$ rather than $b$
it is convenient to consider the complex plane of the variable
$\beta$ where $\beta=b^2$ and analyze singularities in  $\beta$--plane.
 Using  (\ref{vrq}) we can write down the
integral over the contour $C$ around a positive axis
in the $\beta$--plane:
\begin{equation}\label{cont}
  F^{*}(W^2,t,Q^2)= -i\frac{W^2}{2\pi^2}\int_C F^*(W^2,\beta,Q^2)K_0(\sqrt{t\beta})d\beta,
\end{equation}
where $K_0$ is the modified Bessel function and the variable $W^2$
was used instead of the variable $s$.
The contour $C$ can be closed at infinity
and the value of the integral will be then
determined by the singularities of the function
$F^*(W^2,\beta,Q^2)$ in the complex $\beta$--plane (Fig. 7), where
\[
F^*(W^2,\beta,Q^2)=\omega(\beta,Q^2)
  \frac{U(W^2,\beta)}{1-iU(W^2,\beta)}.
\]

\begin{figure}[htb]
\vspace{2mm}
\hspace{3cm}
\epsfxsize=3.1in \epsfysize=1.85in
 \epsffile{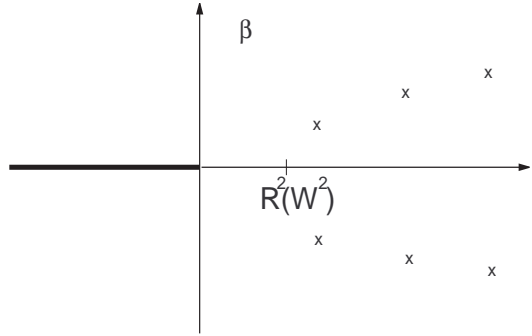}
 \caption[impl]{The singularities of the scattering amplitude in the complex
 $\beta$-plane.}
\label{fig:xx}
\end{figure}

With the explicit expressions for the functions $U$ and $\omega$ we conclude
that the positions of the poles which are determined by  solutions
of the equation
\[
1-iU(W^2,\beta)=0
\]
are located at
\[
\beta_n(W^2)=[R(W^2)+ i\frac{\xi}{M}\pi n]^2, \quad n=\pm 1,\pm 3,\ldots\; .
\]
where $M=\tilde{m}_Q n_V+m_Qn_h$ and
\[
R(W^2)=\frac{\xi}{M}\ln \left[g^N\left (\frac{W^2}{{ m_Q}^2}
\right )^{N/2}\right].
\]
 The location of the poles does not depend on the virtuality $Q^2$.

Besides the poles the function $F^*(W^2,\beta,Q^2)$ has a branching point at $\beta=0$
and
\[
\mbox{disc}\; F^*(W^2,\beta,Q^2)=
\]
\[
\frac{\mbox{disc} [\omega(\beta,Q^2)U(W^2,\beta)]-iU(W^2,\beta+i0)U(W^2,\beta-i0)
\mbox{disc}\; \omega(\beta ,Q^2)}{[1-iU(W^2,\beta + i0)][1-iU(W^2,\beta -i0)]},
\]
i.e.
\[
\mbox{disc}\; F^*(W^2,\beta,Q^2)\simeq i\;\mbox{disc}\; \omega(\beta,Q^2)
\]
since at  $W^2\to\infty$ the function $U(W^2,\beta)\to\infty$ at fixed $\beta$.
As a result the function $F^{*}(W^2,t,Q^2)$  can be represented as a sum of pole
and cut contributions, i.e.
\[
F^{*}(W^2,t,Q^2)=F_p^{*}(W^2,t,Q^2)+F_c^{*}(W^2,t,Q^2).
\]
The pole and
cut contributions are decoupled dynamically when $W^2\rightarrow
\infty  $. Contribution of
the poles determines the
amplitude  $F^{*}(W^2,t,Q^2)$ in the region $|t|/W^2 \ll 1$
and it can be written in a form of  series:
\begin{equation}\label{polc}
F^*(W^2,t,Q^2)\simeq iW^2(W^2)^{\lambda_V(Q^2)/2}\sum_{n=\pm 1,\pm 3,\ldots}
\exp\left\{\frac{i\pi n}{N}\lambda_V(Q^2)\right\}\sqrt{\beta_n}
K_0(\sqrt{t\beta_n}).
\end{equation}
 At moderate values of $-t$ when $-t \geq 1$ (GeV/c)$^2$ the
  amplitude (\ref{polc}) leads to
the Orear type behavior of the differential cross--section which is similar to
the Eq.(\ref{ore}) for the on--shell amplitude, i.e.
\begin{equation}\label{orev}
  \frac{d\sigma_V}{dt}\propto\exp\left[-\frac{2\pi\xi}{M}\sqrt{-t}\right].
\end{equation}

Note that at small $t$ the behavior of the differential cross--section
 is complicated.  The oscillating factors
$\exp\left\{\frac{i\pi n}{N}\lambda_V(Q^2)\right\}$
 absent in the on-shell scattering amplitude \cite{csn}, play a role.

At large $t$ the poles contributions is negligible  and contribution
from the cut at $\beta=0$ is a dominating one.
It appears that   the function $F_c^{*}(W^2,t,Q^2)$ does not depend  on energy
and differential cross section depends on $t$ in a power-like way
\begin{equation}\label{ttri}
  \frac{d\sigma_V}{dt}\simeq \tilde{G}(Q^2)\left(1-\frac{\bar{\xi}^2(Q^2)t}{\tilde{m}_Q^2}
  \right)^{-3}.
\end{equation}
Therefore for large values of $-t$
($-t\gg \tilde{m}_Q^2/\bar{\xi}^2(Q^2)$)
we have a simple $(-t)^{-3}$ dependence of the differential cross--section.
This dependence is very distinct  from the corresponding behavior
of the differential cross--section
of the on-shell
scattering \cite{csn} which approximates the quark counting rule \cite{matv}
 because due to the off-shell unitarity effects.
It is worth noting that the
 ratio of the two differential cross-sections for the production of the
vector mesons $V_1$ and $V_2$
  does
not depend on the variables $W^2$ and $t$ at large  values of $t$.

\section{Impact parameter picture}
The results described above rely on the off--shell unitarity
and the $Q^2$--dependence of constituent quark interaction radius.
It is useful to consider an impact parameter picture to get
insight into the physical origin of this $Q^2$--dependence.
An impact parameter analysis of the experimental data
was a particular tool for the detection of
the unitarity effects in hadronic reactions \cite{amal}
 and, as it was proposed in \cite{pred}, similar technique can be used
 in the diffractive DIS.
Impact parameter profile of the amplitude
 is peripheral when $\xi(Q^2)$
increases with $Q^2$ (Fig. 8). The dependence on virtuality
of constituent quark interaction radius was assumed
 and this dependence
appeared to be in a qualitative agreement with the experimental data.
 It was demonstrated then that the rising dependence of the constituent quark
interaction radius with virtuality
 implies the rising $Q^2$-dependence of the exponent $\lambda(Q^2)$.
The relation  between $\xi(Q^2)$ and $\lambda(Q^2)$
 implies in its turn a saturation of the $Q^2$-dependence of $\lambda(Q^2)$ at large
values of $Q^2$.
\begin{figure}[htb]
\vspace{2mm}
\hspace{4.5cm}
\epsfxsize=3.1in \epsfysize=1.85in
 \epsffile{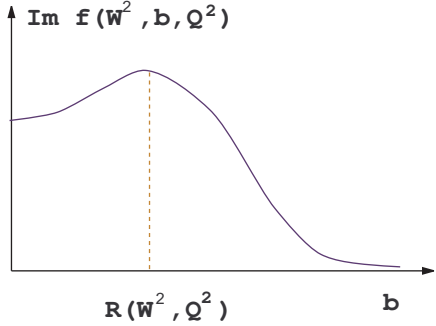}
 \caption[imp]{ The impact parameter profile of the scattering amplitude.}
\label{fig:2}
\end{figure}
The reason
 for the increase of the constituent quark interaction radius with
 virtuality
should have a dynamical nature and it could originate from
 the emission of the additional $q\bar q$--pairs in the
 nonperturbative  structure of a constituent quark.
In the present
approach constituent quark consists of a current quark
and the  cloud of quark--antiquark pairs of the different
flavors \cite{csn}.
It was shown that the available experimental data
imply  $\ln Q^2$--dependence for the  radius of this
cloud.

The peripheral profile of the amplitude in its turn can result from the
increasing role of the orbital angular momentum of the quark--antiquark cloud
when the virtual particles are considered. The generation of
 $\bar q q$-pairs cloud  could be considered in analogy with the theory
 of superconductivity.
It was proposed \cite{spcon} to push further this analogy and consider
 an  anisotropic extension of the theory of superconductivity
 which seems to match well with the above nonperturbative
  picture for a constituent
  quark.  The studies \cite{anders} of that theory show that the
   presence of anisotropy leads to axial symmetry of pairing
  correlations around the anisotropy direction $\hat{\vec{l}}$ and to
 the particle currents induced by the pairing correlations.  In
 another words it means that a particle of the condensed fluid
   is surrounded
 by a cloud of correlated particles (``hump'')  which rotate around
it with the
axis of rotation $\hat {\vec l}$.
 Calculation of the
orbital momentum  shows that it is proportional to the density of the
correlated particles.
The value of the orbital
 momentum contribution into the spin of constituent quark can be
 estimated according to  the relation between
 contributions of current quarks into a proton spin and corresponding
contributions of current quarks into a spin of the constituent quarks
and that of the constituent quarks into  proton spin.

It is evident that the results of the impact parameter analysis are in
favor
of the increasing role of the orbital angular momenta  with virtuality.

\section{Conclusion}

We considered limitations the unitarity provides for the $\gamma^* p$--total
 cross-sections and geometrical effects in the
  model dependence of $\sigma^{tot}_{\gamma^* p}$.
In particular, it was shown that the $Q^2$--dependent constituent quark
 interaction radius    can lead to a nontrivial,
asymptotical result:
$\sigma^{tot}_{\gamma^* p}\sim (W^2)^{\lambda(Q^2)}$,
where $\lambda(Q^2)$ will be saturated at large values of $Q^2$.
This result is valid  when the interaction radius of the virtual
constituent quark is rising with the virtuality $Q^2$.
The  data for the structure functions
  at low values of $x$ continue to demonstrate the
  rising total cross-section of $\gamma^* p$--interactions
 and therefore we can consider it as a reflection of
  the rising with virtuality
  interaction radius of a constituent quark.
Thus, we have shown that the power-like parameterization of the experimental
data $\sigma^{tot}_{\gamma^* p}\sim (W^2)^{\lambda(Q^2)}$ with $Q^2$--dependent
exponent can have a physical ground and should not  be regarded merely as
a convenient way to represent the data.  Other scenarios
which are consistent with unitarity have also been  discussed.
General conclusion is the following: unitarity itself does not lead
to the saturation at $x\to 0$, i.e.  slow down of the
power-like energy dependence of $\sigma^{tot}_{\gamma^* p}$ and transition
to the energy behavior  consistent with the Froissart--Martin bound valid
for the on--shell  scattering.

In  elastic vector meson electroproduction processes
the mass and $Q^2$ dependencies of the integral
cross--section  of vector meson production
are related to the  dependence of the interaction radius  of the constituent
quark $Q$  on the respective quark mass $m_Q$ and its virtuality
$Q^2$.
The behavior of the differential cross--sections at large $t$ is in large
extent determined by the off-shell unitarity effects. The smooth power-like
dependence on $t$ is predicted.
New experimental data would
have an essential meaning for  discrimination
 of the model approaches and studies of the
interplay between the non-perturbative and perturbative QCD regimes.

\section*{Acknowledgements} We are grateful to J. A. Crittenden  for
 the communications on the ZEUS experimental data on the angular distributions
 and A. Borissov for the discussions of the HERMES experimental data.
 We would also  like to thank
 M. Islam, E. Martynov, V. Petrov
 and A. Prokudin for  many interesting discussions of the results.


\small
\begin{thebibliography}{99}
\bibitem{her}
A. M. Cooper-Sarkar, R. C. E. Devenish and A. De Roeck,
Int. J. Mod. Phys. A \bf13\rm,  3385 (1998).
\bibitem{indur}
C. Lopez and F. J. Yndurain, Phys. Rev. Lett. \bf44\rm,   1118 (1980).
\bibitem{pqcd}
 V. N. Gribov and L. N. Lipatov, Sov. J. Phys. \bf15\rm,  438, 625 (1972); \\
L. N. Lipatov,  Sov. J. Nucl. Phys. \bf20\rm, 94 (1975);\\
Yu. L. Dokshitzer, Sov. Phys. JETP \bf46\rm,  641 (1977);\\
G. Altarelli and G. Parisi, Nucl. Phys. B \bf426\rm,  298 (1977).
\bibitem{lipa} L. N. Lipatov, Sov. J. Nucl. Phys. \bf23\rm,  338 (1976);\\
E. A. Kuraev, L. N. Lipatov and V. S. Fadin, Sov. Phys. JETP \bf45\rm,  199 (1977);\\
Y. Y. Balitsky and L. N. Lipatov, Sov. J. Nucl. Phys. \bf28\rm,   822 (1978).
\bibitem{nad}
P. M. Nadolsky, S. M. Troshin and N. E. Tyurin, Z. Phys. C \bf69\rm,  131 (1995).
\bibitem{petr}
V. A. Petrov, Nucl. Phys. Proc. Suppl. \bf54A\rm,  160 (1997);\\
V. A. Petrov and A. V. Prokudin, Proceedings of the International
Conference on Elastic and Diffractive Scattering, Protvino, Russia,
28 June - 2 July 1999, p.95, World Scientific, 2000,
V. A. Petrov and A. V. Prokudin, eds.
\bibitem{lands}
P. V. Landshoff, hep-ph/0010315;\\
A. Donnachie, J. Gravelis and G. Shaw, hep-ph/0101221;\\
S. Munier,  A. M. Sta\'sto and A. H. Mueller, hep-ph/0102291.
\bibitem{h1} C. Adloff at al. [H1 Collaboration], Preprint  DESY 01-104, 2001.
\bibitem{mart}
 P. Desgrolard, A. Lengyel, E. Martynov, hep-ph/0110149.
\bibitem{wolf}
G. Wolf, hep-ex/0105055.
\bibitem{ttpre}
S. M. Troshin and N. E. Tyurin, Europhys. Lett. \bf37\rm,   239 (1997).
\bibitem{levin}
A. L. Ayala, M. B. Gay Ducati and E. M. Levin, Phys. Lett. B \bf388\rm, 188 (1996);\\
A. Capella, E. G. Ferreiro,  A. B. Kaidalov and C. A.Salvado, Nucl. Phys.  B \bf 593
\rm, 336 (2001).
\bibitem{pas}
E. A. Paschos,
Phys. Lett.  B \bf389\rm,  383 (1996);\\
W. L. van Neerven,
Nucl. Phys. B, Proc. Suppl. \bf79\rm,  36 (1999).
\bibitem{csn} S. M. Troshin and N. E.Tyurin,
 Nuovo Cim. \bf106A\rm,  327 (1993); Phys. Rev. D \bf49\rm,  4427 (1994).
\bibitem{blan}
R. Blankenbecler and M. L. Goldberger. Phys. Rev. \bf 126\rm , 766 (1962).
\bibitem{log}
A. A. Logunov, M. A. Mestvirishvili, Nguen Van Hieu and O. A. Khrustalev,
Nucl. Phys. B \bf10\rm,  692 (1969).
\bibitem{chy}
 T. T. Chou and C. N. Yang, Phys. Rev. \bf170\rm,   1591 (1968).
\bibitem{khru}
O. A. Khrustalev, V. I. Savrin and N. E. Tyurin, Comm. JINR E2-4479, 1969.
\bibitem{bart}
A. M. Sta\'sto, K. Golec-Biernat and J. Kwieci\'nski, hep-ph/0007192;\\
J. Bartels and H. Kowalski, hep-ph/0010345.
\bibitem{pred}
B. Povh, B. Z. Kopeliovich and E. Predazzi, Phys. Lett. \bf B405\rm , 361 (1997).
\bibitem{zosa}
ZEUS Collaboration, J.~Breitweg { et al.}, Paper 439 submitted to
the XXXth Interanational Conference on High Energy
Physics, July27 - August 2, 2000, Osaka, Japan.
\bibitem{melld}
R. Ioshida, hep-ph/0102262 and references therein.
\bibitem{lov}
C. Lovelace, Phys. Rev. \bf135\rm, B 1225 (1964).
\bibitem{zdta}
 ZEUS Collaboration, J.~Breitweg { et al.}, Paper 442 submitted to
the XXX International Conference on High Energy Physics, 27 July - 2
August, 2000, Osaka, Japan.
\bibitem{critt}
J. A. Crittenden, hep-ex/0010079, references therein and private communication.
\bibitem{lang}
S. M. Troshin and N. E. Tyurin, Theor. Math. Phys. \bf50\rm, 150 (1982).
\bibitem{cald}
A. C. Caldwell and M. S. Soares, hep-ph/0101085.
\bibitem{ivan}
D. Yu. Ivanov, Phys. Rev. \bf D53\rm , 3564;\\ D. Yu. Ivanov, R. Kirschner,
 A. Scha\"afer and L. Szymanowski, Phys. Lett, \bf B478\rm , 101, 2000.
\bibitem{forsh}
J. R. Forshaw and G. Poludniowski,
hep-ph/0107068.
\bibitem{gots}
E. Gotsman, E. Levin, U. Maor and E. Naftali, TAUP 2691/2001, hep-ph/0110256.
\bibitem{diehl}
M. Diehl, hep-ph/0109040.
\bibitem{matv}
V. A. Matveev, R. M. Muradyan and A. N. Tavkhelidze, Lett. Nuovo.
Cim. \bf7\rm,  719 (1973);\\
S. Brodsky and G. Farrar, Phys. Rev. Lett. \bf31\rm,  1153 (1973).
\bibitem{amal}
U. Amaldi and K. R. Schubert, Nucl. Phys. \bf B166\rm , 301 (1980).
\bibitem{spcon} S. M.
Troshin and N. E.  Tyurin, Phys. Rev.  \bf D52\rm,   3862, (1995);
ibid. \bf D54\rm,   838, (1996);
  Phys. Lett.  \bf B355\rm ,
 543, (1995).
\bibitem{anders} P. W. Anderson and P. Morel, Phys. Rev. \bf
123\rm, 1911,  (1961);\\ F. Gaitan, Annals of Phys. \bf
235\rm,  390, (1994);\\ G. E. Volovik, Pisma v ZhETF,
 \bf 61\rm, 935, (1995).


\end{thebibliography}
\end{document}